\documentclass[astrosymb,twocolumn,trackchanges]{aastex631}

\usepackage{amsmath,amssymb}
\usepackage[caption=false]{subfig}
\usepackage{chngpage}
\renewcommand\AA{\mathring{\text{A}}}

\received{December 5, 2024}

\submitjournal{ApJL}

\begin{document}

\title{Lack of Rest-frame UV Variability in Little Red Dots Based on HST and JWST Observations}

\author[0000-0003-0747-1780]{Wei Leong Tee}
\affiliation{Steward Observatory, University of Arizona, 
933 N Cherry Ave, Tucson, AZ 85719, USA }

\author[0000-0003-3310-0131]{Xiaohui Fan}
\affiliation{Steward Observatory, University of Arizona, 
933 N Cherry Ave, Tucson, AZ 85719, USA }

\author[0000-0002-7633-431X]{Feige Wang}
\affiliation{Steward Observatory, University of Arizona, 933 N Cherry Ave, Tucson, AZ 85719, USA }
\affiliation{Department of Astronomy, The University of Michigan, 1085 S. University, 323 West Hall, Ann Arbor, MI 48109-1107, USA}

\author[0000-0001-5287-4242]{Jinyi Yang}
\affiliation{Steward Observatory, University of Arizona, 933 N Cherry Ave, Tucson, AZ 85719, USA }
\affiliation{Department of Astronomy, The University of Michigan, 1085 S. University, 323 West Hall, Ann Arbor, MI 48109-1107, USA}



\begin{abstract}
Variability is a fundamental signature for active galactic nuclei (AGN) activity, and serve as an unbiased indicator for rapid instability happened near the center supermassive black hole (BH).
Previous studies showed that AGN variability does not have strong redshift evolution, and scales with their bolometric luminosity and BH mass, making it a powerful probe to identify low-mass, low-luminosity AGNs at high redshift. 
JWST has discovered a new population of high-redshift galaxies likely hosting  moderate accreting BHs ($> 10^6 M_\odot$) --- the little red dots (LRDs, $z\sim4-10$). 
In this paper, we study the variability of a sample of 21 LRDs with V-shaped SEDs in three JWST deep fields that also have reliable HST observations in closely paired filters at $1-2 \mu$m (rest-frame UV), with the time difference between 6 and 11 years. This LRD sample covers a redshift range of $3 < z < 8$ with $-21.3 < M_{UV} <-18.4$.
Based on both photometry and imaging difference analyses, we find a mean magnitude difference of $\sim0.12\pm0.24$ mag, with none of the LRDs showing photometric variability at 3$\sigma$ significance. 
Extrapolation of SDSS quasar variability predicts a magnitude change of order 0.3 mag for our LRD sample. 
This suggests an upper limit of about $\sim 30\%$ AGN contribution to the total observed UV light in our sample of LRDs. 
\end{abstract}

\keywords{Active galactic nuclei (16), Supermassive black holes (1663), Quasars (1319)}


\section{Introduction}\label{sec:introduction}
Early JWST observations have discovered 
a class of compact galaxies showing ``V-shaped'' spectral energy distribution (SED) with blue UV excess and rapidly raising red optical-to-IR continuum --- usually referred to as the Little Red Dots (LRDS), with strong indication of AGN activities, especially for a subset of that exhibit rest-frame optical broad-line (BL) emissions \citep[$>1000$ kms$^{-1}$, $M_\mathrm{BH}=10^{6-8} \, M_\odot$, $L_\mathrm{bol}\sim 10^{46}\, \mathrm{ergs}^{-1}$,][]{Harikane2023ApJ...959...39H,Kocevski2023ApJ...954L...4K,Labbe2023arXiv230607320L,Maiolino2023arXiv230801230M,Greene2024ApJ...964...39G,Matthee2024ApJ...963..129M}. 
Their clear BL detection and compact red IR continuum emission are suggestive of rapid accreting SMBH activity, however their blue UV spectra are more consistent with star-forming galaxies \citep{Killi2023arXiv231203065K}, with potential small contribution from AGN-scattered light \citep{Matthee2024ApJ...963..129M,Greene2024ApJ...964...39G}. 
In addition, absent of strong detection in LRDs' X-ray stacking analysis \citep{Yue2024arXiv240413290Y} has questioned their AGN origin. 
Although several theories have been proposed in attempt to explain the AGN signatures \citep[e.g.,][]{Maiolino2024arXiv240500504M,Li2024arXiv240710760L}, majority of LRDs have blue UV slope that is hardly consistent with purely AGN radiation, with reported AGN flux fraction ranging from 40\% to 80\% through morphology and SED determinations \citep{Harikane2023ApJ...959...39H,Durodola2024arXiv240610329D}. 
An independent test will further constrain the boundary of AGN contribution.

AGNs are known to vary in brightness ($>$0.1 mag) with timescales ranging from days to years. 
It is thought to originate from the instability in the accretion disk close to the SMBH, therefore is a hall mark of AGN phenomena. 
Extensive research have been conducted for low-redshift quasar/AGN variability in X-ray, ultraviolet (UV) and optical wavelengths \cite[e.\,g.,\,][]{Turner1999ApJ...524..667T,VandenBerk2004ApJ...601..692V,MacLeod2012ApJ...753..106M}. 
The structure function, a power spectrum density measurement of AGN magnitude change over time, is well-described with a time-series damped-random walk (DRW) model, with a characteristic variability amplitude SF$_\infty$ and turn-over time scale $\tau$ \citep[e.g.,][]{MacLeod2012ApJ...753..106M,Kozlowski2016ApJ...826..118K}. 
Given the anti-correlation of rest UV/optical luminosity with variability amplitude \citep{Simm2016A&A...585A.129S}, rest-frame UV-optical variability become a powerful tool to discover and study high-redshift faint AGNs, which may hold a key into constraining SMBH seeding models \citep{Inayoshi2024arXiv240214706I,Jeon2024arXiv240218773J}, the AGN contribution to reionization at cosmic dawn \citep{Finkelstein2019ApJ...879...36F,Dayal2020MNRAS.495.3065D,Yung2021MNRAS.508.2706Y}, and the early coevolution of galaxies and SMBHs \citep{Habouzit2022MNRAS.511.3751H,Inayoshi2022ApJ...927..237I,Pacucci2023ApJ...957L...3P}.
If the rest-frame UV radiation of LRDs is dominated by AGN emission, they are expected to show strong variability at this wavelength.

However, the variability study of high-redshift, low-luminosity AGNs is challenging due to a combination of time dilation and their faintness in the rest-frame UV, which is now redshifted into the red optical and NIR wavelengths. 
\cite{Sanchez2017ApJ...849..110S} reported $10-20\, \%$ NIR variable sources in 1.4 $\mathrm{deg}^2$ COSMOS are AGNs ($M_{Y} \leq -22$ at $z=1-3$). Similar study at $z>4$ are not feasible before JWST launch.  
JWST and HST have similar pairs of filters at $1-2,\mu$m, allowing for the direct study of photometric variations between new JWST observations and previous HST observations of the same field over a time span of 6-11 years. This makes them ideal for searching for variable AGNs at $z>3$.
As an example, \cite{O'Brien2024ApJS..272...19O} searched for transients using two epochs high resolution HST imaging in the JWST Time Domain Field (TDF), and found a dozen of brown dwarfs and $z\lesssim 3$ variable AGN candidates.  \cite{Hayes2024arXiv240316138H} conducted high-redshift transients search in HUDF using two epochs HST photometric data and successfully identified two intermediate redshift AGNs and two reionization era quasar candidates (but see \cite{DeCoursey2024arXiv240605060D} which suggests that one of them is likely a late type star). 

In this paper, we study the rest-frame UV variability of a sample of LRDs identified in the JWST deep field observations with existing HST observations in closely matched filter pairs. 
We first introduce the methodology of the variable source finder in Section \ref{sec:methodology}. We describe the use of public available images/catalogs, data preparation and the LRD sample definition in Section \ref{sec:data}. In Section \ref{sec:result_discussion}, we present the constraints on photometric variabilities of LRDs and its implication on the AGN nature of LRDs. We use AB magnitude throughout this work. Errors are quoted within 68\% confidence interval, and assume a flat $\Lambda$CDM cosmology with $H_\mathrm{0}=70$ km s$^{-1}$ Mpc$^{-1}$, $\Omega_\mathrm{m}=0.3$, $\Omega_\Lambda=0.7$.

\section{Methodology}\label{sec:methodology}
\begin{deluxetable*}{wc{3em}wc{2.5em}wc{2.5em}wc{2.5em}wc{2.5em}wc{3em}wc{3em}wc{5em}wc{10em}wc{20em}}
\tablenum{1}
\tablecaption{Field information where LRDs are selected.}
\label{tab:field}
\tablewidth{0pt}
\tabletypesize{\scriptsize}
\tablehead{& \colhead{5$\sigma$ Depth ($d\sim 0.3\arcsec$)} \span\span\span\span  & & \colhead{Overlapped} & & \\
\colhead{Field} & 
\begin{tabular}{c}
   JWST\\F115W
\end{tabular} & 
\begin{tabular}{c}
   HST\\F125W
\end{tabular} & 
\begin{tabular}{c}
   JWST\\F150W
\end{tabular} & 
\begin{tabular}{c}
   HST\\F160W
\end{tabular} & 
$\overline{T}_\mathrm{HST}$ & $\overline{T}_\mathrm{JWST}$ &
\begin{tabular}{c}
     Area \\
     (arcmin$^2$)
\end{tabular} & 
\begin{tabular}{c}
     Related \\
     Surveys
\end{tabular} & 
\colhead{References}
}
\startdata
EGS & 27.3 &27.0 &27.0 & 27.1& 2012.14& 2022.82&188 & 
\begin{tabular}{c}
   HST/CANDELS\\
   JWST/CEERS
\end{tabular} & 
\begin{tabular}{c}
 \cite{Grogin2011ApJS..197...35G,Koekemoer2011ApJS..197...36K}\\
 \cite{Bagley2023ApJ...946L..12B}, Tee et al. (in prep.)
 \end{tabular}\\
GOODS-S &29.3 &27.9 &29.3 &27.6 & 2011.29& 2023.19&159 & 
\begin{tabular}{c}
   HST/HLF\\
   JWST/JADES
\end{tabular} & 
\begin{tabular}{c}
\cite{Illingworth2016arXiv160600841I,Whitaker2019ApJS..244...16W}\\
 \cite{Eisenstein2023arXiv231012340E}, Tee et al. (in prep.)
 \end{tabular}\\
Abell 2744 & 28.2& 27.1& 28.2& 26.5& 2016.03&2023.21 &148 & 
\begin{tabular}{c}
   HST/BUFFALO, HFF\\
   JWST/UNCOVER, MegaScience
\end{tabular}
& 
\begin{tabular}{c}
     \cite{Weaver2024ApJS..270....7W}\\
      \cite{,Suess2024arXiv240413132S}
\end{tabular}
\enddata
\end{deluxetable*}

\subsection{Imaging Data and Photometry}\label{sec:data}
We study the variability of LRDs selected from three deep fields with both HST and JWST observations: EGS, GOOD-S and Abell 2744. The basic properties of these imaging dataset are summarized in Table \ref{tab:field}. 
In order to carry out consistent photometric measurements and image subtractions, all image mosaics have been registered to Gaia reference frame, and the HST mosaics have been resampled to common tangential plane as with the JWST image and common pixel scales of $0.\arcsec03$/pixel 
For EGS and GOODS-S images, we do PSF-matching for JWST images to HST/F160W because the latter has the broadest full-width-half-maximum (FWHM) in $1-2\,\mu$m, photometry extraction and estimation are done on low-resolution HST and JWST images. More detail regarding to data preparation can be found in Tee et al. (in preparation).

\subsection{Photometry Calibration}
Both HST and JWST have similar filter coverage over $1-2\,\mu$m, i.e., F125W and F160W with HST's Wide Field Camera 3 (WFC3; $5\sigma \,\sim 26-27$ mag), 
F115W and F150W by JWST's Near Infrared Camera (NIRCam; $5\sigma \,\sim 28-29$ mag). 
Given the fact that JWST/NIRCam imaging depth is in general 2 mag deeper than HST/WFC3, we can use the higher quality JWST magnitudes to color calibrate the HST magnitudes and account for the small filter differences. 
We choose a simple color correction model for HST magnitude using JWST magnitude and color, 
\begin{equation}
    m_\mathrm{HST} = \beta + m_\mathrm{JWST} + \alpha \times (m_\mathrm{F115W}-m_\mathrm{F150W})
    \label{eq:hst_jwst_calibration}
\end{equation}
The calibrated HST photometry, $m_\mathrm{HST_\mathrm{exp}}$, namely \textit{expected HST photometry based on JWST observations}, should compared with measured value $m_\mathrm{HST}$. 
We choose to adopt aperture photometry for the calibration since our goal is to study variable AGNs, in which the brightness should only change at the galactic centers. If a source is detected in JWST filters but become significantly brighter or fainter in HST filters, it will be classified as potential variable AGN. 
We use a 0.25\arcsec aperture radius (0.24\arcsec in UNCOVER catalog) 
for aperture flux estimation, about 2 times larger than HST/F160W point-spread-function (PSF) full-width-half-maximum (FWHM; $\sim 0.\arcsec 22$). The photometric uncertainties are estimated by randomly placing apertures within survey areas masking out detected sources, and we add a noise floor of 0.01 for magnitude systematic errors.
We construct a sample of sources with $m=22-28$, compare their observed HST magnitudes with JWST-calibrated HST magnitudes, and find negligible systematic offsets.

\subsection{Variable Source Classification}
We define the photometric variability in the flux space, $\mathrm{var}_p$, as 
\begin{equation}
    \mathrm{var}_{p,b}=\frac{f_\mathrm{exp}-f}{\sqrt{{\sigma_{\mathrm{exp}}^2} + {\sigma}^2}}
    \label{eq:photo_var}
\end{equation}
where $p$ stands for photometry method, $b$=HST/F125W,HST/F160W. 
A source is considered variable if $ \mathrm{var}_{p,b}>3$.

Pure aperture photometry selection may include contamination from supernovae which are off-centered, or moving transients. We also construct a difference image to validate the nature of variability. 
We create 20$\arcsec$ $\times$ 20$\arcsec$ JWST/F115W, HST/F125W, JWST/F150W, HST/F160W cutouts for each source, 
and scale the JWST cutout flux level to HST cutout. 
After preparation of the background subtracted photometry and astrometry aligned cutouts, we use the PyZOGY\footnote{https://github.com/dguevel/PyZOGY} prescription \citep{Zackay2016ApJ...830...27Z}, which calculate the difference image $D$ given input pair images and corresponding PSFs, i.e. (JWST/F115W, HST/F125W) and (JWST/F150W,HST/F160W). We measure the flux difference $\Delta f_D$
within aperture directly from difference image $D$, and evaluate the imaging difference variability, $\mathrm{var}_D$, as 
\begin{equation}
    \mathrm{var}_{D,b} = \frac{\Delta f_{D,b}}{\sqrt{{\sigma_{\mathrm{exp}}^2} + {\sigma}^2}}
    \label{eq:im_var}
\end{equation}
where $b$=HST/F125W,F160W, assuming no other dominant noise component present. \cite{Zackay2016ApJ...830...27Z} showed that $D$ image is a proper difference image, where the noise level is dominated by white noise only given reasonable fit to the flux zero points in pair images. 
We use both $\mathrm{var}_{p,b}$ and $\mathrm{var}_{D,b}$ to justify the variability nature of variable sources.

\begin{figure}[ht]
    \centering
    \includegraphics[width=\linewidth]{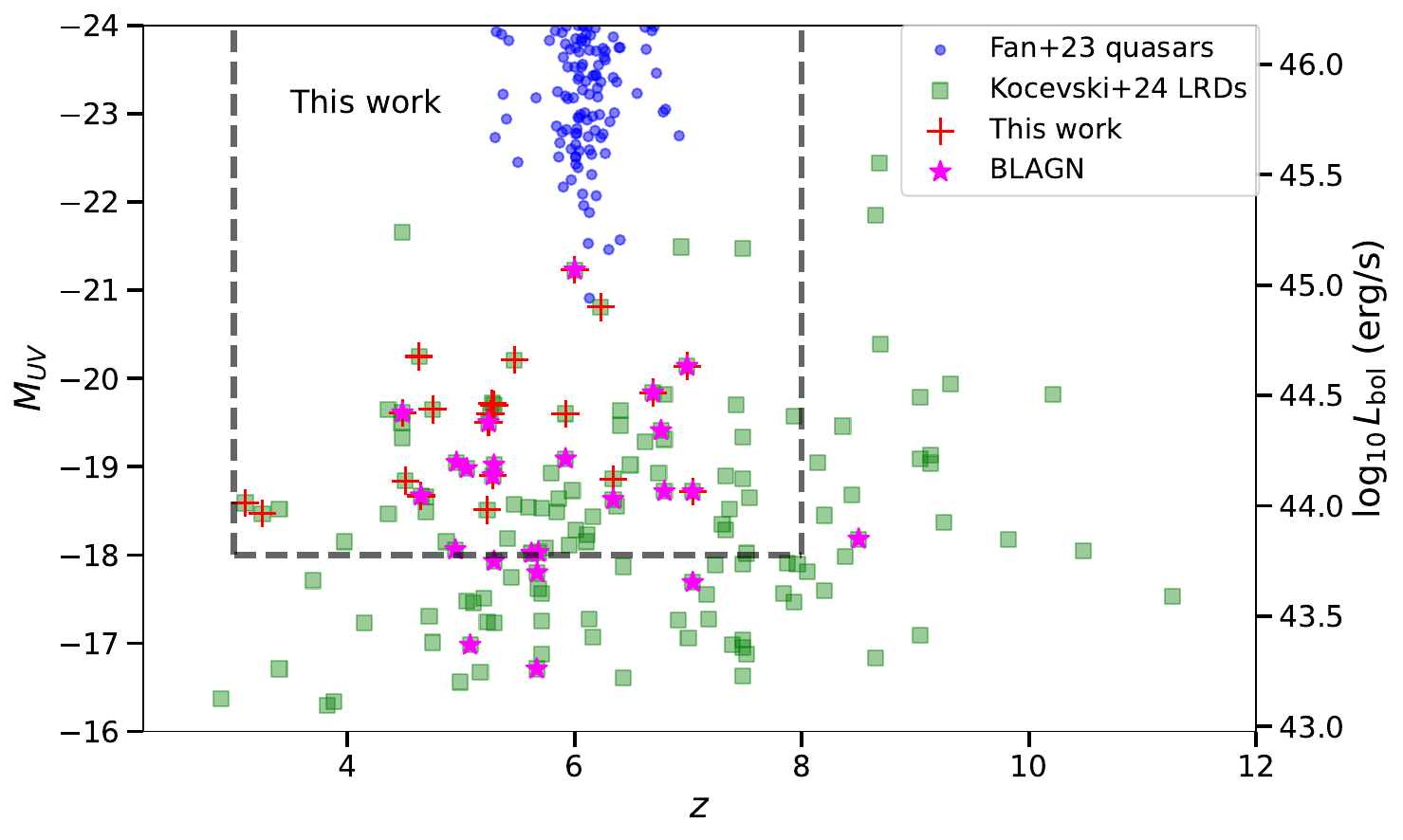}
    \caption{Redshift and luminosity distribution of LRDs in the parent sample \citep[green square, $N=143$,][]{Kocevski2024arXiv240403576K}, broad-line AGN subsample (magenta stars), $z>5.5$ quasars \citep[blue circle,][]{Fan2023ARA&A..61..373F}. 
    We construct an unbiased flux-limited sample ($M_\mathrm{UV}<-18$) and show the 21 LRDs (8 BLAGNs) considered in this work with red pluses. }
    \label{fig:z_M}
\end{figure}

\subsection{LRD Sample Definition}\label{sec:lrd_def}
\cite{Kokorev2024ApJ...968...38K} constructed a sample of LRDs at $z=4-9$, selected by their JWST colors and compact sizes. 
\cite{Kocevski2024arXiv240403576K} further characterized the LRDs selection by including $\beta_\mathrm{UV}$ slope and $\beta_\mathrm{opt}$ to select primarily ``V-shaped" SED, with or without broad Balmer line detections in publicly available spectroscopic data. 
They reported a total of 341 $z\sim 2-11$ LRDs in JWST public extragalactic surveys, including CEERS (EGS), PRIMER (UDS and COSMOS), JADES (GOODS-S/N), UNCOVER (Abell 2744) and NGDEEP (HUDF) programs. In this work, we focus our study on NIRCam short wavelength (SW) data to characterize the high-redshift LRDs rest-frame near-UV variability, using mainly CEERS, JADES and UNCOVER data where public data are officially released.
The data preparation detail on EGS and GOODS-S imaging and catalog products will be addressed in Tee et al. (in preparation). For Abell 2744 we use the images and catalog generated by UNCOVER team for variability analysis\citep{Bezanson2024ApJ...974...92B,Weaver2024ApJS..270....7W,Suess2024arXiv240413132S}. 

The median JWST/F444W magnitude of LRDs is about 26 mag, but they are fainter in rest-frame UV to about 28.5 mag, which is close to the detection limit in JWST/F115W and JWST/F150W, 1-2 mag deeper than the depth typically of the avaialble HST/F125W and HST/F160W imaging. 
We use the following three selection criteria to create a sample of LRDs suitable for variability studies: 
1. The source is $5\sigma$ detected in JWST/F115W and JWST/F150W, and has mutual image coverage in HST/F125W and HST/F160W. 
2.  The color-calibrated $f_\mathrm{exp}$ in the corresponding HST observations is 3 times above HST noise $\sigma$. This way, if the source is not varying, it have been detected in either epoch. Thus, the selection is not biased against high variable sources that are not detected in HST imaging as a result of their variability. 
3. We limit the redshift $z<8$ to avoid Ly$\alpha$ drop out affecting the calibration between JWST/F115S and HST/F125W bands. 
The initial catalog consists of 143 LRDs in three fields, which later reduces to 21 sources (14\% of original sample) for a flux-limited sample spanning $-21.23<M_\mathrm{UV}<-18.47$. This sample includes 8 BLAGNs (1 without BH mass measurements) identified in literature \citep{Harikane2023ApJ...959...39H,Kocevski2024arXiv240403576K,Maiolino2023arXiv230801230M,Greene2024ApJ...964...39G}. 
We extract their photometry in low-resolution PSF-matched images, and calculate their expected HST fluxes and flux errors. Next we perform imaging difference on photometrically and astrometrically aligned images. Finally we derive $ \mathrm{var}_{p} $ by calculating the significance between expected flux and measured flux, $\mathrm{var}_{D} $ by measuring the flux residual significance in difference image $D$.

\section{Result and Discussion}\label{sec:result_discussion}
\begin{figure}[ht]
    \centering
    \includegraphics[width=\linewidth]{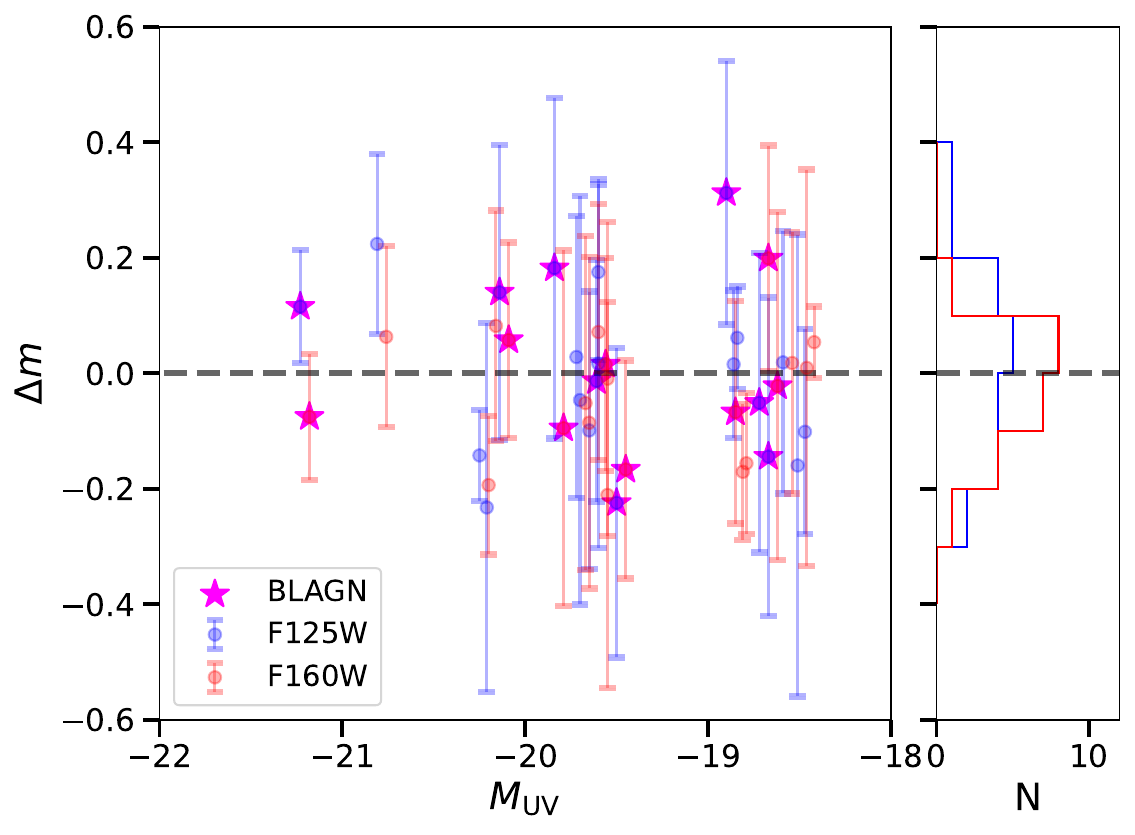}
    \caption{Magnitude change between HST and JWST epoch of time difference 6-11 years. The mean magnitude changes in HST/F125W and HST/F160W are $\sim 0.12\pm0.24$ mag. }
    \label{fig:delta_m}
\end{figure}

The redshift and luminosity distribution of all LRDs in the parent sample are displayed in Fig. \ref{fig:z_M}, along with $z>5.5$ faint quasars \citep{Fan2023ARA&A..61..373F}. We highlight the 21 LRDs/8 BLAGN  in this study. 
Their photometry and variability measurements are reported in Table \ref{tab:variability_result}. 
Majority of the sources show $\lesssim0.2$ magnitude change, as shown in Fig. \ref{fig:delta_m}. The mean magnitude changes in HST/F125W and HST/F160W are 0.12$\pm$0.24 and 0.09$\pm$0.22 mag respectively.
Several sources show $>0.2$ magnitudes, noticeably 
CEERS\_1670, CEERS\_1669, CEERS\_5208, CEERS\_13318, CEERS\_24253 and UNCOVER\_9497. 
We calculate their magnitude errors through quadratic sum, $\mathrm{err}_{\Delta m} = \sqrt{\mathrm{err}^2_\mathrm{obs} + \mathrm{err}^2_\mathrm{exp}} $, where individual magnitude error is derived from empirical noise function measured by placing fixed radius apertures across random position in the image (Tee et al., in prep.). 
We find that he magnitude errors are dominated by shallower HST observation, and none show $>3\sigma$ variability.

\begin{figure}[ht]
    \centering
    \includegraphics[width=\linewidth]{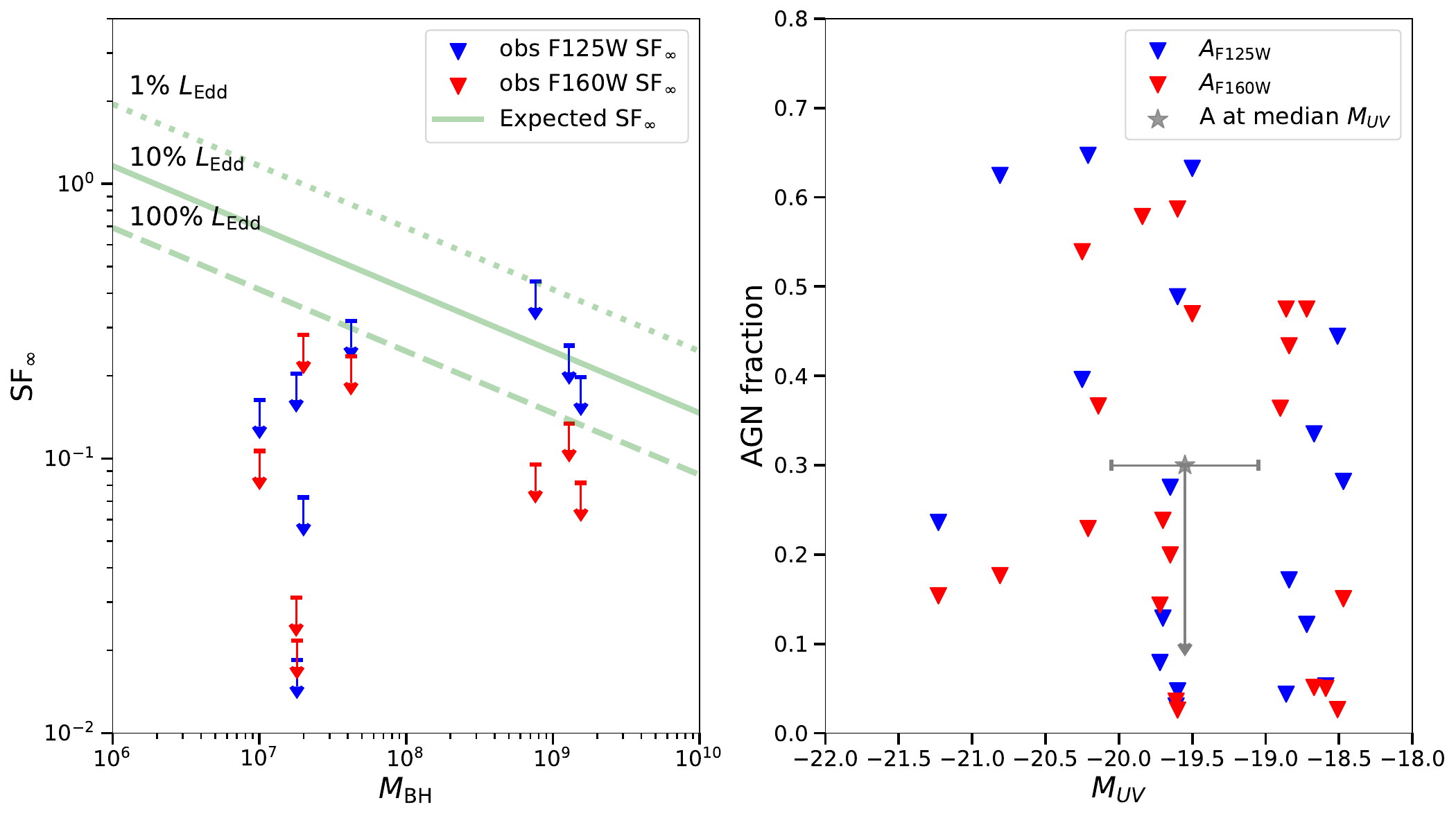}
    \caption{SF$_\infty$ as a function of $M_\mathrm{BH}$ (left) and AGN fraction as function of $M_{UV}$ (right). The asymptotic variability amplitude is derived from the maximum magnitude change between the HST and JWST epochs in HST/F125W and HST/F160W. The green line illustrates the asymptotic variability amplitude extrapolated from SDSS quasar relation from \cite{MacLeod2010ApJ...721.1014M}. The AGN fraction is a first order estimation of the AGN flux contribution to the magnitude change, $A=\frac{\mathrm{SF}_{\infty,\mathrm{d}}}{\mathrm{SF}_{\infty}}$, the mean of 21 LRDs results in an upper limit of 30\% UV flux are AGN dominated.}
    \label{fig:sf_inf}
\end{figure}

The LRDs in this work are selected via V-shaped SED, among them eight also  show broad H$\alpha$ lines and hence classified as BLAGN. \cite{Labbe2023arXiv230607320L} shows that SEDs of LRDs suggest a bolometric luminosities $L_\mathrm{bol}=10^{43-46}$ erg s$^{-1}$, or equivalently $M_\mathrm{BH} = 10^{7-9} M_\odot$ \citep{Greene2024ApJ...964...39G}, representing a class of low-medium luminosity AGN hosting overmassive BH at the centers. 
\cite{Kokubo2024arXiv240704777K} reported no variability detected for 5 broad H$\alpha$ emitters in GLASS survey, having a median $L_\mathrm{bol}=10^{44}$ erg s$^{-1}$, and typical BH mass of $M_\mathrm{BH}\sim 10^8 M_\odot$, a typical LRD sample. 
To constraint the origin of rest-frame UV emission of LRDs, we compare their asymptotic variability amplitude SF$_\infty$, with the expected value derived from literature studies of variable AGN light curves. 
Observational SF$_\infty$ is calculated via SF$_\infty = \sqrt{2}\sqrt{(m_2 - m_1)^2-\sigma^2_\mathrm{non-var}}$ \citep{MacLeod2010ApJ...721.1014M}, where $m_1,m_2$ are two epoch magnitudes, $\sigma_\mathrm{non-var}$ is the magnitude scatter of non-variable objects in the field with similar time difference. 
Here we drop the magnitude scatter term, and simply compute the upper limit of SF$_\infty$.
\cite{MacLeod2010ApJ...721.1014M} has first determined the dependencies of optical SF$_\infty$ on several AGN properties, including AGN luminosity, redshift, $M_\mathrm{BH}$ etc, through modeling the SDSS Stripe 82 quasar light curves as DRW (also see \citealt{Kozlowski2016ApJ...826..118K}). 
Structure function approaches the SF$_\infty$ when the time difference between two epochs is greater than the characteristic turnover timescale $\tau$, i.e. entering the white noise regime, which is found to be approximately several hundred days \citep{Kozlowski2016ApJ...826..118K},  \citep{Burke2021Sci...373..789B}. 
Based on the median $6-11$ years observed time differences between past HST observation and the recent JWST observation, we assume that SF$_\infty$ derived are indeed upper limit of the structure function.
We adopt the following relation to scale with the rest-frame wavelength ($\lambda_\mathrm{RF}$) and AGN luminosity ($M_i$) from \cite{MacLeod2010ApJ...721.1014M} (without $M_\mathrm{BH}$ dependence, seebelow), 
\begin{equation}
      \log \mathrm{SF}_\infty = -0.618 - 0.479 \log \left(\frac{\lambda_\mathrm{RF}}{4000 \AA}\right)+0.09 (M_i + 23) 
\label{eq:sf_infty}
\end{equation}
where $M_i$ is the absolute $i$ band magnitude $K$-corrected to $z=2$, which is representative of AGN bolometric luminosity ($L_\mathrm{bol}$) as shown in \cite{Richards2006AJ....131.2766R}). 
Here we have two options to compute $M_i$: 1. Adopt the $M_{UV}$ computed in \cite{Kocevski2024arXiv240403576K} using rest-frame 1450$\AA$ and convert that to $M_i$ using Eq. 3 in \cite{Richards2006AJ....131.2766R}. 2. Following \cite{Shen2009ApJ...697.1656S}, $M_i$ can be characterized with $M_i = 90-2.5 \log L_\mathrm{bol}/\mathrm{erg\,s^{-1}}$, and assuming a fixed $L_\mathrm{bol}/L_\mathrm{Edd}$, $M_i$ is a $M_\mathrm{BH}$ dependent function. The first option assumes a full contribution of UV light from AGN, while the second option is an extrapolation to low-mass AGN from SDSS quasars \citep{Burke2023MNRAS.518.1880B}. 
Note that in our sample most of the LRDs do not have $M_\mathrm{BH}$ information, except for the known BLAGNs (Fig. \ref{fig:sf_inf}).
We choose to calculate expected SF$_\infty$ with LRDs median $M_\mathrm{BH}$ of $10^{7.6} M_\odot$ within our sample when $M_\mathrm{BH}$ are not available, and a fixed $L_\mathrm{bol}/L_\mathrm{Edd}=0.1$. We find the expected SF$_\infty$ is 0.3-0.6 mag.

Note that first method postulates a  model in which full UV light is emitted by AGN \citep[see discussion in ][]{Greene2024ApJ...964...39G},
while the second method assumes a $M_\mathrm{BH}$ dependent function extrapolated from SDSS Stripe 82 quasars. 
Fainter, smaller LRDs are subjected to have more host galaxy light dilution, results in smaller SF$_\infty$ \citep{Burke2023MNRAS.518.1880B}.
Studies on host diluted SF$_\infty$ have been conducted extensively, i.e. \cite{Burke2023MNRAS.518.1880B} (and reference therein). In order to obtain a first order estimation on host galaxy flux contribution, we assume the small flux variation is entirely by AGN and $\Delta f_\mathrm{AGN} \ll f_\mathrm{AGN}$ and the fact that SF $\propto \frac{\Delta f}{f}$. We define the AGN fraction $A$,
\begin{equation}
     A = \frac{f_\mathrm{AGN}}{f_\mathrm{AGN}+f_\mathrm{host}} = \frac{\mathrm{SF}_{\infty,\mathrm{d}}}{\mathrm{SF}_{\infty}}
\end{equation}
where $\mathrm{SF}_{\infty,\mathrm{d}}$ and $\mathrm{SF}_{\infty}$ are the host-diluted and normal variability amplitudes respectively. In our case observational $\mathrm{SF}_{\infty}$ are treated as $\mathrm{SF}_{\infty,\mathrm{d}}$. 

Following the two options to calculate AGN UV light fraction, we compute the mean of AGN fraction of the 21 sources in our sample. 
SDSS quasars have $M_\mathrm{BH}$  that are  1-2 dex more massive.
The fact that SF$_{\infty}$ scales inversely with $M_\mathrm{BH}$ ($\log \mathrm{SF}_\infty \propto M_i \propto -\log L_\mathrm{bol} \propto M_\mathrm{BH}$) results in a lower limit if extrapolated from SDSS variability amplitude relation, which we interpret as an upper limit in AGN fraction. 
Using $M_\mathrm{BH}=10^{7.6} M_\odot$ and $M_{UV}=-19.6$,
we find that  the average upper limit of AGN fraction in our sample is $\sim$30\%. 
In other words, AGN can contribute at most 1/3 of the total fluxes in LRD rest-frame UV. We show the overall result in Fig. \ref{fig:sf_inf}.
Our result is in general lower than those reported in literature with SED analysis on faint AGNs and LRDs at similar redshift range. \cite{Harikane2023ApJS..265....5H} reported a contribution of $\sim 50\%$ using faint BLAGNs, while \cite{Durodola2024arXiv240610329D} measured $40-85\%$ AGN contribution in LRDs spectra.
Clearly a larger sample with multi-epoch deep JWST data will benefit the variability studies and provide better constrain on the AGN origin of LRDs rest-UV spectra. 
The upcoming JWST GO programs, e.g., COSMOS-3D and NEXUS, provide excellent opportunity to investigate the LRDs' variability soon. 
COSMOS-3D (GO-5893) will provide one additional epoch of JWST/F115W imaging on top of COSMOS-Web (GO-1727) footprint in COSMOS field; NEXUS (GO-5105) will map around North Ecliptic Pole (NEP) with wavelength coverage $0.9-12\,\mu$m with three epochs. Both surveys will contribute a total of $\sim0.4 \, \mathrm{deg}^2$ area, the largest JWST multi-epoch datasets existed for variable AGN studies. 
The methodology of this study will extend in the future to include those surveys.

\section{Conclusion}
In this study, we constrain the fraction of AGN contribution in LRDs using rest-UV variability. Our sample consists of 30 bright LRDs selected from the public JWST extragalactic field surveys, photometric selected through V-shaped SEDs.
Our study use both HST and JWST observations in the same field coverage, taking advantages of long time span to measure the variability of LRDs. 
We find a $\sim 0.12\pm 0.24$ magnitude change spanning $6-11$ years in the rest-frame UV. We find no source with $3\sigma$ variability using both direct photometry and imaging difference techniques.
We derive the expected variability amplitude of LRDs by extrapolation from SDSS-like luminous AGNs, and estimate the AGN fraction by comparing the observed variability amplitude to prediction. Our result suggests that AGN contributing $\lesssim 30\%$ in the total UV light. Our results are consistent with the scenario that the blue UV continuum is not dominated by the AGN contributions. 

\begin{acknowledgments}
W.L.T appreciate the comments and suggestions from J. Lyu, J. Helton, F. Sun, Y. Wu, Z. Ji, Z. Chen, X. Jin, C. Burke, J. Champagne, M. Pudoka, W. Liu, H. Zhang, C. DeCoursey, X. Lin, Y. Zhu, M. Rieke 
for useful and informative discussion. W.L.T. appreciate G. Hosseinzadeh help on imaging difference. 
W.L.T and X.F. acknowledge support from HST-AR-17565.
F.W. acknowledge support from NSF Grant AST-2308258.
W.L.T. would like to thank the support by Lia YCC throughout the work.
\end{acknowledgments}

%

\vspace{5mm}


\software{Astropy, Numpy, Scipy, Photutils, PyZOGY}

\bibliography{var_agn_LRD}{}
\bibliographystyle{aasjournal}
\begin{deluxetable*}{ccccccccccc}
\rotate
\tablenum{2}
\tablecaption{Table on LRDs photometry and variability.}
\label{tab:variability_result}
\tablehead{
\colhead{ID} & \colhead{RA} &\colhead{DEC} &\colhead{$z_\mathrm{best}$} &\colhead{$M_\mathrm{UV}$} &\colhead{$\log_\mathrm{10} (M_\mathrm{BH})$} & \colhead{$m_\mathrm{F150W}$} &\colhead{$\Delta m_\mathrm{F125W}$} & \colhead{$\Delta m_\mathrm{F160W}$}  &  \colhead{BL} &
\colhead{BL and $M_\mathrm{BH}$ Ref.}  \\
& & & & mag & $M_\odot$ & mag & mag & mag &  Detected &  
}
\startdata
JADES\_8083	& 03:32:31.88 & -27:48:06.70 & 4.65 & -18.67 & 7.25 & 27.68 $\pm$ 0.04 & -0.14 $\pm$ 0.28 & -0.02 $\pm$ 0.3 & Y &  \cite{Maiolino2023arXiv230801230M}  \\
JADES\_4454	& 03:32:39.87 & -27:46:19.35 & 6.34 & -18.86 & - & 28.07 $\pm$ 0.07 & 0.02 $\pm$ 0.13 & -0.17 $\pm$ 0.12 & N &    \\
JADES\_11125 & 03:32:16.24 & -27:48:44.39 & 4.63 & -20.25 & - & 26.28 $\pm$ 0.03 & -0.14 $\pm$ 0.08 & -0.19 $\pm$ 0.12 & N &   \\
JADES\_20532 & 03:32:29.94 & -27:51:58.68 & 4.51 & -18.84 & - & 26.27 $\pm$ 0.03 & 0.06 $\pm$ 0.09 & -0.16 $\pm$ 0.12 & N &   \\
JADES\_21925 & 03:32:20.84 & -27:52:23.00 & 3.1 & -18.59 & - & 27.34 $\pm$ 0.03 & 0.02 $\pm$ 0.23 & 0.02 $\pm$ 0.23 & N &  \\
JADES\_24052 & 03:32:17.79 & -27:53:01.33 &  5.23 & -18.51 & -  & 27.43 $\pm$ 0.09 & -0.16 $\pm$  0.4 & 0.01 $\pm$ 0.34 &  N  & \\
JADES\_26901 & 03:32:23.41 & -27:54:04.52 & 5.26 & -19.6 & - & 26.44 $\pm$ 0.04 & 0.18 $\pm$ 0.15 & -0.01 $\pm$ 0.27 & N &  \\
CEERS\_397 & 14:19:20.69 & +52:52:57.70 & 6.0 & -21.23 & 7.0 & 26.03 $\pm$ 0.04 & 0.12 $\pm$ 0.1 & -0.08 $\pm$ 0.11 & Y & \cite{Harikane2023ApJ...959...39H}   \\
CEERS\_1236 & 14:20:34.87 & +52:58:02.20 & 4.484 & -19.61 & 7.26 & 26.91 $\pm$ 0.07 & -0.01 $\pm$ 0.21 & 0.02 $\pm$ 0.18 & Y & \cite{Harikane2023ApJ...959...39H}  \\
CEERS\_1465 & 14:19:33.12 & +52:53:17.70 & 5.274 & -19.72 & - & 27.09 $\pm$ 0.09 & 0.03 $\pm$ 0.24 & -0.05 $\pm$ 0.29 & N & \\
CEERS\_1670 & 14:19:17.63 & +52:49:49.01 & 5.242 & -19.5 & 7.62 & 26.63 $\pm$ 0.07 & -0.22 $\pm$ 0.27 & -0.17 $\pm$ 0.19 & Y & \cite{Harikane2023ApJ...959...39H}  \\
CEERS\_1669 & 14:19:16.25 & +52:52:40.28  &  5.47 & -20.21 & - & 26.88 $\pm$  0.08 &  -0.23 $\pm$  0.32  &  0.08 $\pm$ 0.2  & N & \\
CEERS\_5208 & 14:19:42.54 & +52:56:02.01  &  5.92 &  -19.6 & - & 27.26 $\pm$ 0.11  &  0.02 $\pm$ 0.32 & -0.21 $\pm$ 0.33 &  N & \\  
CEERS\_7902 & 14:19:55.93 & +52:57:21.62 & 6.99 & -20.14 & 9.19 & 26.71 $\pm$ 0.07 & 0.14 $\pm$ 0.26 & 0.06 $\pm$ 0.17 & Y & \cite{Kocevski2024arXiv240403576K} \\
CEERS\_10444 & 14:19:34.14 & +52:52:38.66  & 6.69 & -19.84 & 9.11 & 27.11 $\pm$ 0.1 & 0.18 $\pm$ 0.3 & -0.09 $\pm$ 0.31 & Y & \cite{Kocevski2024arXiv240403576K}\\ 
CEERS\_13318 & 14:19:10.89 & +52:47:19.85 & 5.28 & -18.9 & 8.88 & 26.93 $\pm$ 0.08 & 0.31 $\pm$ 0.23 & -0.07 $\pm$ 0.19 & Y & \cite{Kocevski2024arXiv240403576K}  \\
CEERS\_19578 & 14:19:31.24 & +52:48:45.22 &  5.29 & -19.7 &- & 27.11 $\pm$ 0.1 &  -0.05 $\pm$ 0.35 &  -0.09 $\pm$  0.29 & N & \\
CEERS\_24253 & 14:19:55.19 & +52:51:39.88 & 6.23 & -20.81 & - & 26.52 $\pm$ 0.06 & 0.22 $\pm$ 0.16 & 0.06 $\pm$ 0.16 & N & \\
UNCOVER\_9497 & 00:14:19.16 & -30:24:05.65 & 7.04 & -18.72 & 7.3 & 27.96 $\pm$ 0.14 & -0.05 $\pm$ 0.26 & 0.2 $\pm$ 0.2 & N & \cite{Greene2024ApJ...964...39G}   \\
UNCOVER\_9447 & 00:14:30.03 & -30:24:05.10 & 3.25 & -18.47 & - & 24.35 $\pm$ 0.02 & -0.1 $\pm$ 0.18 & 0.05 $\pm$ 0.06 & N &   \\
UNCOVER\_3943 & 00:13:59.72 & -30:21:10.60 & 4.75 & -19.65 & - & 25.98 $\pm$ 0.03 & -0.1 $\pm$ 0.24 & 0.07 $\pm$ 0.22 & N & 
\enddata
\end{deluxetable*}
\end{document}